# Lattice Boltzmann modeling of multiphase flows at large density ratio with an improved pseudopotential model

Q. Li,[1] K. H. Luo,[1,2] and X. J. Li[3]

[1]Energy Technology Research Group, Faculty of Engineering and the Environment, University of Southampton, Southampton SO17 1BJ, United Kingdom

[2]Center for Combustion Energy, Key Laboratory for Thermal Science and Power Engineering of Ministry of Education, Tsinghua University, Beijing 100084, China

[3]School of Civil Engineering and Mechanics, Xiangtan University, Xiangtan 411105, China

Owing to its conceptual simplicity and computational efficiency, the pseudopotential multiphase lattice Boltzmann (LB) model has attracted significant attention since its emergence. In this work, we aim to extend the pseudopotential LB model to simulate multiphase flows at large density ratio and relatively high Reynolds number. First, based on our recent work [Li *et al*., Phys. Rev. E. **86**, 016709 (2012)], an improved forcing scheme is proposed for the multiple-relaxation-time pseudopotential LB model in order to achieve thermodynamic consistency and large density ratio in the model. Next, through investigating the effects of the parameter $a$ in the Carnahan-Starling equation of state, we find that the interface thickness is approximately proportional to $1/\sqrt{a}$. Using a smaller $a$ will lead to a wider interface thickness, which can reduce the spurious currents and enhance the numerical stability of the pseudopotential model at large density ratio. Furthermore, it is found that a lower liquid viscosity can be gained in the pseudopotential model by increasing the kinematic viscosity ratio between the vapor and liquid phases. The improved pseudopotential LB model is numerically validated via the simulations of stationary droplet and droplet oscillation. Using the improved model as well as the above treatments, numerical simulations of droplet splashing on a thin liquid film are conducted at a density ratio in excess of 500 with Reynolds numbers ranging from 40 to 1000. The dynamics of droplet splashing is correctly reproduced and the predicted spread radius is found to obey the power law reported in the literature.

## I. INTRODUCTION

The lattice Boltzmann (LB) method [1-4], which originates from the lattice gas automaton (LGA) method, has been developed into an alternative numerical approach for simulating fluid flows and solving nonlinear problems. Different from the conventional numerical methods, the LB method is based on the mesoscopic kinetic equation for particle distribution functions. Because of its kinetic nature, the LB method has been found to be particularly useful in modeling interfacial phenomena in multiphase flows [4-6]. Among the existing multiphase LB models [7-17], the pseudopotential LB model proposed by Shan and Chen [10, 11] has received considerable attention for its simplicity and computational efficiency. An attractive feature of the pseudopotential LB model is that the phase segregation can emerge naturally as a result of particle interactions, without tracking or capturing the interface between different phases, which is required in many other numerical approaches.

However, the pseudopotential LB model also suffers from several severe drawbacks, such as large spurious currents and thermodynamic inconsistency [16], and is usually limited to low-density-ratio problems. For the sake of overcoming these drawbacks, numerous studies have been conducted from both theoretical and numerical viewpoints. Shan [18] proposed to reduce the spurious currents using high-order isotropic discrete gradient operators. Sbragaglia *et al*. [19] developed a multirange pseudopotential model by combining the nearest-neighbor interactions and the next-nearest-neighbor interactions. In addition, Sbragaglia *et al*. [20] have devised a free-energy formulation of the pseudopotential LB model. Furthermore, several attempts have been made by Yuan and Schaefer [21], Falcucci *et al*. [22], and Kupershtokh *et al*. [23] to break through the low-density-ratio restriction.

Despite the fact that great efforts have been made, modeling multiphase flows at large density ratio and high Reynolds number using the pseudopotential LB model is still very challenging because many issues should be addressed simultaneously, e.g., how to realize large density ratio, how to eliminate thermodynamic inconsistency, how to reduce spurious currents, and how to achieve high Reynolds number (low viscosity). For some of the above-mentioned techniques, as reported by Huang *et al*. [24], their achievable largest density ratio will drop rapidly when the viscosity decreases. Meanwhile, most previous studies of the pseudopotential LB model are focused on stationary/quasi-stationary multiphase problems.

In the LB community, there have been several LB models for simulating large-density-ratio multiphase flows, such as the multiphase LB models devised by Zheng *et al*. [25] and Lee *et al*. [17]. Nevertheless, Zheng *et al*.'s model was found to be restricted to density-matched binary fluids and unable to simulate multiphase flows with noticeable density differences [26]. Lee *et al*.'s model has successfully reproduced some multiphase flows at large density ratio, but the numerical algorithm is complex and its computational efficiency may be low because the numerical implementation involves the discretization of many derivatives. By employing a multiple-relaxation-time (MRT) collision operator in Lee *et al*.'s model, Mukherjee and Abraham [27] have constructed an incompressible MRT multiphase LB model. In addition, McCracken and Abraham [28] have also proposed an incompressible MRT multiphase LB model based on He *et al*.'s multiphase LB model [29]. Besides these models, a MRT free-energy LB model has been devised by Pooley *et al*. [30, 31] and a MRT pseudopotential LB model has been formulated by Yu and Zhao [32]. Using the MRT free-energy model, Pooley *et al*. [30] have accurately reproduced the well-known Washburn's law and they found that the MRT model can stop the unphysical currents appearing near to the interfaces for simulating wetting dynamics [31]. Generally, the MRT models are better than the standard BGK models [33] in terms of numerical stability. It should be noted that, however, the numerical stability of BGK models can also be enhanced via the entropic LB approach [34, 35].

In the present work, we aim to extend the pseudopotential LB model to simulate multiphase flows at large density ratio and relatively high Reynolds number in consideration of its distinct advantages. The MRT collision model is adopted. In order to resolve the problem of thermodynamic inconsistency and realize large density ratio in the MRT pseudopotential LB model, an improved forcing scheme will be proposed based on our recent finding [36] that the thermodynamic consistency can be approximately achieved in the pseudopotential LB model through adjusting the mechanical stability condition. Moreover, the influences of the parameter $a$ in the Carnahan-Starling equation of state will be investigated, and it will be shown that the interface thickness is approximately proportional to $1/\sqrt{a}$ and thus can be widened with a smaller $a$. With the increase of the interface thickness, the spurious currents can be reduced and the numerical stability can be enhanced. Furthermore, it will be shown that a lower liquid viscosity can be obtained in the pseudopotential LB model with the increase of the kinematic viscosity ratio between the vapor and liquid phases. Using these strategies, numerical simulations will be carried out for two-dimensional droplet splashing on a thin liquid film at a density ratio larger than 500 with Reynolds numbers ranging from 40 to 1000.

The rest of the present paper is organized as follows. Section II will briefly introduce the MRT pseudopotential LB model. In Sec. III, an improved forcing scheme will be proposed. Numerical investigations will be presented in Sec. IV. Finally, a brief conclusion will be given in Sec. V.

## II. MRT PSEUDOPOTENTIAL LB MODEL

With the MRT collision operator [37, 38], the evolution equation of the density distribution function can be written as [28, 32, 39]

$$f_\alpha\left(\mathbf{x}+\mathbf{e}_\alpha\delta_t, t+\delta_t\right) = f_\alpha(\mathbf{x},t) - \overline{\Lambda}_{\alpha\beta}\left(f_\beta - f_\beta^{eq}\right)\Big|_{(\mathbf{x},t)} + \delta_t\left(S_\alpha - 0.5\overline{\Lambda}_{\alpha\beta}S_\beta\right)\Big|_{(\mathbf{x},t)}, \tag{1}$$

where $f_\alpha$ is the density distribution function, $f_\alpha^{eq}$ is its equilibrium distribution, $t$ is the time, $\mathbf{x}$ is the spatial position, $\mathbf{e}_\alpha$ is the discrete velocity along the $\alpha$ th direction, $\delta_t$ is the time step, $S_\alpha$ is the forcing term in the velocity space, and $\overline{\Lambda} = \mathbf{M}^{-1}\Lambda\mathbf{M}$ is the collision matrix, in which $\mathbf{M}$ is an orthogonal transformation matrix and $\Lambda$ is a diagonal Matrix given by (for the D2Q9 lattice)

$$\Lambda = \mathrm{diag}\left(\tau_\rho^{-1}, \tau_e^{-1}, \tau_\varsigma^{-1}, \tau_j^{-1}, \tau_q^{-1}, \tau_j^{-1}, \tau_q^{-1}, \tau_\upsilon^{-1}, \tau_\upsilon^{-1}\right). \tag{2}$$

Through the transformation matrix $\mathbf{M}$, the density distribution function $f_\alpha$ and its equilibrium distribution $f_\alpha^{eq}$ can be projected onto the moment space via $\mathbf{m} = \mathbf{M}\mathbf{f}$ and $\mathbf{m}^{eq} = \mathbf{M}\mathbf{f}^{eq}$, respectively. For the D2Q9 lattice, the equilibria $\mathbf{m}^{eq}$ can be given by

$$\mathbf{m}^{eq} = \rho\left(1, -2+3|\mathbf{v}|^2, 1-3|\mathbf{v}|^2, v_x, -v_x, v_y, -v_y, v_x^2-v_y^2, v_xv_y\right)^{\mathrm{T}}. \tag{3}$$

With Eqs. (2) and (3), the right-hand side of the MRT LB equation (1) can be rewritten as [39]

$$\mathbf{m}^* = \mathbf{m} - \Lambda\left(\mathbf{m} - \mathbf{m}^{eq}\right) + \delta_t\left(\mathbf{I} - \frac{\Lambda}{2}\right)\overline{\mathbf{S}}, \tag{4}$$

where $\mathbf{I}$ is the unit tensor and $\overline{\mathbf{S}} = \mathbf{M}\mathbf{S}$ is the forcing term in the moment space, in which $\mathbf{S} = \left(S_0, S_1, \cdots, S_8\right)^{\mathrm{T}}$. The streaming process is given by

$$f_\alpha\left(\mathbf{x}+\mathbf{e}_\alpha\delta_t, t+\delta_t\right) = f_\alpha^*(\mathbf{x},t), \tag{5}$$

where $\mathbf{f}^* = \mathbf{M}^{-1}\mathbf{m}^*$. The corresponding macroscopic density and velocity are calculated by

$$\rho = \sum_\alpha f_\alpha, \quad \rho\mathbf{v} = \sum_\alpha \mathbf{e}_\alpha f_\alpha + \frac{\delta_t}{2}\mathbf{F}, \tag{6}$$

where $\mathbf{F} = \left(F_x, F_y\right)$ is the force acting on the system.

In the pseudopotential LB model, the interaction force, which is used to mimic the molecular interactions that cause phase separation, is given by [18, 40]

$$\mathbf{F} = -G\psi(\mathbf{x}) \sum_{\alpha=1}^{N} w\left(|\mathbf{e}_\alpha|^2\right) \psi(\mathbf{x}+\mathbf{e}_\alpha) \mathbf{e}_\alpha, \tag{7}$$

where $\psi(\mathbf{x})$ is the interaction potential, $G$ is the interaction strength, and $w\left(|\mathbf{e}_\alpha|^2\right)$ are the weights. For the case of nearest-neighbor interactions on the D2Q9 lattice, the weights $w\left(|e_\alpha|^2\right)$ are $w(1) = 1/3$ and $w(2) = 1/12$. Through the Taylor expansion, the leading terms of the interaction force can be obtained [19, 40]

$$\mathbf{F} = -Gc^2 \left[ \psi \nabla \psi + \frac{1}{6} c^2 \psi \nabla \left(\nabla^2 \psi\right) + \cdots \right], \tag{8}$$

where $c$ is the lattice constant. Usually, in the MRT LB method the force is incorporated via the following forcing scheme [28, 32]

$$\bar{\mathbf{S}} = \begin{bmatrix} 0 \\ 6(v_x F_x + v_y F_y) \\ -6(v_x F_x + v_y F_y) \\ F_x \\ -F_x \\ F_y \\ -F_y \\ 2(v_x F_x - v_y F_y) \\ (v_x F_y + v_y F_x) \end{bmatrix}. \tag{9}$$

Actually, Eq. (9) can be treated as the MRT version of Guo *et al.*'s forcing scheme [41], which is widely used in the BGK LB method. Using the Chapman-Enskog analysis, the following Navier-Stokes equations can be derived from Eqs. (4), (5), (6), and (9) in the low Mach number limit:

$$\partial_t \rho + \nabla \cdot (\rho \mathbf{v}) = 0, \tag{10a}$$

$$\partial_t (\rho \mathbf{v}) + \nabla \cdot (\rho \mathbf{v v}) = -\nabla \cdot (\rho c_s^2 \mathbf{I}) + \nabla \cdot \mathbf{\Pi} + \mathbf{F}, \tag{10b}$$

where $c_s = c/\sqrt{3}$ is the sound speed and $\mathbf{\Pi} = \rho \upsilon \left[ \nabla \mathbf{u} + (\nabla \mathbf{u})^T \right] + \rho(\xi - \upsilon)(\nabla \cdot \mathbf{u})\mathbf{I}$ is the viscous stress tensor, in which $\upsilon = c_s^2 (\tau_\upsilon - 0.5)\delta_t$ is the kinematic viscosity and $\xi = c_s^2 (\tau_e - 0.5)\delta_t$ is the bulk viscosity.

## III. IMPROVED FORCING SCHEME

In this section, an improved forcing scheme will be devised for the MRT pseudopotential LB model. To start with, the mechanical stability condition of the pseudopotential LB model and the problem of thermodynamic inconsistency are introduced for general readers.

### A. Mechanical stability condition and thermodynamic inconsistency

According to Eqs. (8) and (10b), the equation of state of the pseudopotential LB model is given by

$$p = \rho c_s^2 + \frac{Gc^2}{2}\psi^2 . \tag{11}$$

Meanwhile from Eq. (10b) the pressure tensor $\mathbf{P}$ can be defined as follows:

$$\nabla \cdot \mathbf{P} = \nabla \cdot \left(\rho c_s^2 \mathbf{I}\right) - \mathbf{F} . \tag{12}$$

With mathematical manipulation, Eq. (8) can be rewritten as

$$\begin{aligned}
\mathbf{F} &= -Gc^2 \left[ \psi \nabla \psi + \frac{c^2}{6}\psi \nabla\left(\nabla^2 \psi\right) \right] + \cdots \\
&= -\frac{Gc^2}{2}\nabla \psi^2 - \frac{Gc^4}{6}\left[\nabla\left(\psi \nabla^2 \psi\right) - \nabla^2 \psi \nabla \psi\right] + \cdots \\
&= -\frac{Gc^2}{2}\nabla \psi^2 - \frac{Gc^4}{6}\nabla\left(\psi \nabla^2 \psi\right) + \frac{Gc^4}{6}\left[\nabla \cdot \left(\nabla \psi \nabla \psi\right) - \frac{1}{2}\nabla|\nabla \psi|^2\right] + \cdots ,
\end{aligned} \tag{13}$$

where $|\nabla \psi|^2 = \left(\partial_x \psi\right)^2 + \left(\partial_y \psi\right)^2$. By combining Eq. (12) and Eq. (13), the continuum form pressure tensor can be obtained

$$\mathbf{P}_c = \left(\rho c_s^2 + \frac{Gc^2}{2}\psi^2 + \frac{Gc^4}{12}|\nabla \psi|^2 + \frac{Gc^4}{6}\psi \nabla^2 \psi\right)\mathbf{I} - \frac{Gc^4}{6}\nabla \psi \nabla \psi + O\left(\partial^4\right). \tag{14}$$

However, Shan [40] argued that, in order to guarantee the exact mechanical balance, the discrete form pressure tensor must be used in the pseudopotential LB model, which can be derived from the volume integral of Eq. (12), i.e.,

$$\int (\nabla \cdot \mathbf{P}) \mathrm{d}\Omega = \int \nabla \cdot (\rho c_s^2 \mathbf{I}) \mathrm{d}\Omega - \int \mathbf{F} \mathrm{d}\Omega, \tag{15}$$

where $\Omega$ is a closed volume. Applying the Gauss integration theorem to Eq. (15) yields

$$\int \mathbf{P} \cdot \mathrm{d}\mathbf{A} = \int \rho c_s^2 \mathbf{I} \cdot \mathrm{d}\mathbf{A} - \int \mathbf{F} \mathrm{d}\Omega, \tag{16}$$

where $\mathrm{d}\mathbf{A}$ is an area element. In discrete form, the above equation becomes

$$\sum \mathbf{P} \cdot \mathbf{A} = \sum \rho c_s^2 \mathbf{I} \cdot \mathbf{A} - \sum \mathbf{F}. \tag{17}$$

According to Eq. (17), the discrete form pressure tensor is defined as [40, 42]

$$\mathbf{P} = \rho c_s^2 \mathbf{I} + \frac{G}{2} \psi(\mathbf{x}) \sum_{\alpha=1}^{N} w(|\mathbf{e}_\alpha|^2) \psi(\mathbf{x} + \mathbf{e}_\alpha) \mathbf{e}_\alpha \mathbf{e}_\alpha. \tag{18}$$

For the case of nearest-neighbor interactions, applying the Taylor expansion to Eq. (18) will yield

$$\mathbf{P} = \left( \rho c_s^2 + \frac{Gc^2}{2} \psi^2 + \frac{Gc^4}{12} \psi \nabla^2 \psi \right) \mathbf{I} + \frac{Gc^4}{6} \psi \nabla \nabla \psi. \tag{19}$$

According to Eq. (19), for a flat interface the normal pressure tensor is given by [40]

$$P_n = \rho c_s^2 + \frac{Gc^2}{2} \psi^2 + \frac{Gc^4}{12} \left[ \alpha \left( \frac{\mathrm{d}\psi}{\mathrm{d}n} \right)^2 + \beta \psi \frac{\mathrm{d}^2 \psi}{\mathrm{d}n^2} \right], \tag{20}$$

where $n$ denotes the normal direction of the interface. For the case of nearest-neighbor interactions, $\alpha$ and $\beta$ are given by $\alpha = 0$ and $\beta = 3$, respectively.

On the basis of Eq. (20) and the requirement that at equilibrium $P_n$ should be equal to the constant static pressure in the bulk [40], the mechanical stability condition can be obtained (see Appendix A for details):

$$\int_{\rho_g}^{\rho_l} \left( p_0 - \rho c_s^2 - \frac{Gc^2}{2} \psi^2 \right) \frac{\psi'}{\psi^{1+\varepsilon}} \mathrm{d}\rho = 0, \tag{21}$$

where $\psi' = \mathrm{d}\psi/\mathrm{d}\rho$, $\varepsilon = -2\alpha/\beta$, and $p_0 = p(\rho_l) = p(\rho_g)$, in which $\rho_l$ is the density of the liquid phase and $\rho_g$ is the density of the vapor phase. In the pseudopotential LB model, the coexistence curves ($\rho_l$ and $\rho_g$) are determined by the mechanical stability condition. However, in the thermodynamic theory the Maxwell equal-area rule which determines the thermodynamic coexistence is built in terms of the following requirement [5]:

$$\int_{\rho_g}^{\rho_l} (p_0 - p_{EOS}) \frac{1}{\rho^2} d\rho = 0. \tag{22}$$

Here $p_{EOS}$ is the equation of state in the thermodynamic theory and $p_0 = p_{EOS}(\rho_l) = p_{EOS}(\rho_g)$. Generally, the mechanical stability condition will lead to different values of liquid and vapor densities in comparison with the solution given by the Maxwell construction. In the pseudopotential LB model, this problem is usually called thermodynamic inconsistency.

In Ref. [42], Sbragaglia and Shan have proposed an interaction potential $\psi$ as follows:

$$\psi(\rho) = \begin{cases} \exp(-1/\rho), & \varepsilon = 0 \\ \left(\dfrac{\rho}{\varepsilon + \rho}\right)^{1/\varepsilon}, & \varepsilon \neq 0 \end{cases}, \tag{23}$$

which gives $\psi'/\psi^{1+\varepsilon} = 1/\rho^2$. With such a choice, Eqs. (21) and (22) will be nearly the same except for the equation of state. To be consistent with the equation of state in the thermodynamic theory, the potential $\psi$ should be chosen as [16, 21]

$$\psi(\rho) = \sqrt{\frac{2(p_{EOS} - \rho c_s^2)}{Gc^2}}. \tag{24}$$

It can be seen that Eq. (23) and Eq. (24) cannot be satisfied at the same time. Note that, when the potential $\psi$ is defined by Eq. (24), $G$ is used to ensure that the whole term inside the square root is positive [21].

### B. Formulation of the improved forcing scheme

To resolve the problem of thermodynamic inconsistency, an alternative approach has been shown in our previous work [36]. The basic idea is that when $\psi$ is defined by Eq. (24), the thermodynamic consistency can be approximately achieved by employing an appropriate $\varepsilon$ in Eq. (21) which can make the mechanical stability solution approximately identical to the solution given by the Maxwell construction. However, $\varepsilon$ is often fixed when the interactions and the corresponding weights are given. For instance, in the case of nearest-neighbor interactions $\varepsilon = 0$, while in the case of nearest- and next-to-nearest-neighbor interactions $\varepsilon = 10/31$ [42].

According to Eqs. (20) and (21), we can see that $\varepsilon$ can be tuned by making the coefficient before the term $(\mathrm{d}\psi/\mathrm{d}n)^2$ adjustable. Meanwhile, it is noticed that $(\mathrm{d}\psi/\mathrm{d}n)^2$ in Eq. (20) is related to two terms in the pressure tensor: $\nabla\psi\nabla\psi$ and $|\nabla\psi|^2 \mathbf{I}$. Hence the coefficient before the term $(\mathrm{d}\psi/\mathrm{d}n)^2$ can be changed by modifying the coefficient in front of either of them.

Within the framework of the BGK pseudopotential LB model, we have presented [36] an improved forcing scheme which can adjust $\varepsilon$ through modifying the coefficient before the term $\nabla\psi\nabla\psi$ in the pressure tensor. A similar scheme can be devised in the MRT pseudopotential LB model, and it will be found that $\bar{S}_1$, $\bar{S}_2$, $\bar{S}_7$, and $\bar{S}_8$ in Eq. (9) need to be changed. Nevertheless, in the present work, by utilizing the feature of the MRT collision operator, we propose a simpler forcing scheme for the MRT pseudopotential LB model as follows:

$$\bar{\mathbf{S}} = \begin{bmatrix} 0 \\ 6(v_x F_x + v_y F_y) + \dfrac{12\sigma |\mathbf{F}|^2}{\psi^2 \delta_t (\tau_e - 0.5)} \\ -6(v_x F_x + v_y F_y) - \dfrac{12\sigma |\mathbf{F}|^2}{\psi^2 \delta_t (\tau_\varsigma - 0.5)} \\ F_x \\ -F_x \\ F_y \\ -F_y \\ 2(v_x F_x - v_y F_y) \\ (v_x F_y + v_y F_x) \end{bmatrix}, \tag{25}$$

where $|\mathbf{F}|^2 = (F_x^2 + F_y^2)$ and $\sigma$ is used to tune $\varepsilon$. According to the Chapman-Enskog analysis [28] as well as Eq. (8), the following Navier-Stokes equation will be obtained:

$$\partial_t (\rho\mathbf{v}) + \nabla \cdot (\rho\mathbf{v}\mathbf{v}) = -\nabla \cdot (\rho c_s^2 \mathbf{I}) + \nabla \cdot \mathbf{\Pi} + \mathbf{F} - 2G^2 c^4 \sigma \nabla \cdot (|\nabla\psi|^2 \mathbf{I}) + O(\partial^5). \tag{26}$$

Hence Eq. (12) can be rewritten as

$$\nabla \cdot \mathbf{P} = \nabla \cdot \left(\rho c_s^2 \mathbf{I}\right) + 2G^2 c^4 \sigma \nabla \cdot \left(\left|\nabla \psi\right|^2 \mathbf{I}\right) - \mathbf{F} + O(\partial^5). \tag{27}$$

Since the second term on the right-hand side of Eq. (27) is a divergence term, then $2G^2 c^4 \sigma \left|\nabla \psi\right|^2 \mathbf{I}$ can be directly absorbed into the pressure tensor (for both continuum and discrete forms):

$$\mathbf{P}_{new} = \mathbf{P}_{original} + 2G^2 c^4 \sigma \left|\nabla \psi\right|^2 \mathbf{I}. \tag{28}$$

As a result, Eq. (20) should be rewritten as

$$P_n = \rho c_s^2 + \frac{Gc^2}{2}\psi^2 + \frac{Gc^4}{12}\left[(\alpha + 24G\sigma)\left(\frac{d\psi}{dn}\right)^2 + \beta\psi\frac{d^2\psi}{dn^2}\right], \tag{29}$$

which leads to $\varepsilon = -2(\alpha + 24G\sigma)/\beta$. Now the mechanical stability condition ($\varepsilon$) is adjustable. For example, when $\sigma = 0.125$ and $G = -1$, $\varepsilon$ will be given by $\varepsilon = 2$ for the case of nearest-neighbor interactions.

Several statements are made about the proposed forcing scheme. First, it can be seen that the basic strategy of the present scheme is to tune the coefficient before the term $\left|\nabla \psi\right|^2 \mathbf{I}$ in the pressure tensor to make the mechanical stability condition adjustable. Second, by comparing Eq. (25) with Eq. (9), we can see that the simple structure of Eq. (9) is retained and only $\bar{S}_1$ and $\bar{S}_2$ are modified. In particular, there is no appreciable increase in computational cost or memory use. Finally, we would also like to point out the proposed scheme is a compromised approach to eliminating the thermodynamic inconsistency of the pseudopotential LB model. On one hand, the scheme is still very simple and the advantages of the pseudopotential LB model are retained. On the other hand, since the mechanical stability solution is fitted to the solution given by the Maxwell construction via $\varepsilon$, only approximate consistency can be obtained.

## IV. NUMERICAL SIMULATIONS

In this section, numerical investigations will be conducted with the improved pseudopotential LB model. Firstly, the improved forcing scheme will be validated via simulations of stationary droplet and droplet oscillation. Subsequently, the influences of the parameter $a$ in the Carnahan-Starling equation of state on the interface thickness and the spurious velocity will be shown. Finally, numerical simulations will be performed for the problem of droplet splashing on a thin liquid film.

### A. Stationary droplet and droplet oscillation

Two tests are considered to validate the improved pseudopotential LB model. The first test is the problem of stationary droplets, which can be used to compare the numerical coexistence curve with the coexistence curve given by the Maxwell construction. In the present study, the Carnahan-Starling (C-S) equation of state is adopted, which is given by [21]

$$p_{\text{EOS}} = \rho RT \frac{1 + b\rho/4 + (b\rho/4)^2 - (b\rho/4)^3}{(1 - b\rho/4)^3} - a\rho^2, \tag{30}$$

where $a = 0.4963 R^2 T_c^2 / p_c$ and $b = 0.18727 RT_c / p_c$. The corresponding critical density $\rho_c$ is given by $\rho_c \approx 0.5218/b$. Following Ref. [21], in our simulations we set $b = 4$, $R = 1$, $c = 1$, and $\delta_t = 1$. With $b = 4$, $\rho_c$ and $a$ would be given by $\rho_c \approx 0.13045$ and $a = 10.601 RT_c$. In previous studies, $a$ is usually set to be 1.0, and then $T_c = a/(10.601R) \approx 0.094$. Here we use $a = 0.5$ and $T_c \approx 0.047$. The effects of the parameter $a$ will be shown in the next section.

A $200 \times 200$ lattice is adopted and a circular droplet with a radius of $r_0 = 50$ is initially placed at the center of the domain with the liquid phase inside the droplet. The periodical boundary conditions are applied in the $x$- and $y$-directions. The density field is initialized as follows [24]:

$$\rho(x, y) = \frac{\rho_l + \rho_g}{2} - \frac{\rho_l - \rho_g}{2} \tanh\left[\frac{2(r - r_0)}{W}\right], \tag{31}$$

where $W = 5$ and $r = \sqrt{(x-x_0)^2 + (y-y_0)^2}$, in which $(x_0, y_0)$ is the central position of the computational domain. For the C-S equation of state used in the present work, $G = -1$ is used. The relaxation times in Eq. (2) are chosen as follows: $\tau_\rho = \tau_j = 1.0$, $\tau_e^{-1} = \tau_\varsigma^{-1} = 1.1$, and $\tau_q^{-1} = 1.1$.

The coexistence curves of the cases $\tau_\upsilon = 0.6$ and $\tau_\upsilon = 0.8$ are shown in Fig. 1. The parameter $\sigma$ in the improved forcing scheme [Eq.(25)] is set to be $\sigma = 0.11$. For comparison, the results obtained with the original forcing scheme are also presented in Fig. 1. From the figure we can see that, for the original forcing scheme, its achievable lowest reduced temperature is around $T/T_c = 0.8$, with the largest density ratios 30.0 and 56.5 for the cases $\tau_\upsilon = 0.6$ and $\tau_\upsilon = 0.8$, respectively. In contrast, it can be seen that the improved forcing scheme works well at $T/T_c = 0.49$, which corresponds to $\rho_l/\rho_g \sim 900$. Moreover, in the vapor branch the results given by the original forcing scheme significantly deviate from the solution of the Maxwell construction, while the results predicted by the improved forcing scheme are in good agreement with those given by the Maxwell construction in both the liquid and vapor branches. In summary, Fig. 1 clearly demonstrates that the proposed forcing scheme is capable of achieving thermodynamic consistency and large density ratio in the MRT pseudopotential LB model.

Another test is the problem of droplet oscillation. In this problem, the droplet is slightly perturbed from its equilibrium circular shape and exhibits oscillatory behavior. According to Lamb [43], the oscillation period for a two-dimensional droplet is given as follows:

$$T_a = 2\pi \left[ n(n^2 - 1) \frac{\vartheta}{\rho_l R_0^3} \right]^{-1/2}, \qquad (32)$$

where $\vartheta$ is the surface tension, $R_0$ is the equilibrium droplet radius, and $n$ denotes the mode of oscillation, which is given by $n = 2$ for an initial elliptic shape [27]. In simulations, a $200 \times 200$ lattice is used. The reduced temperature is set to be $T/T_c = 0.5$, which corresponds to $\rho_l/\rho_g \sim 700$.

The elliptic droplet is positioned at the center of the computational domain, with the major radius $R_{max} = 30$ and the minor radius $R_{min} = 27$. The equilibrium droplet radius $R_0$ is given by $R_0 = \sqrt{R_{max} R_{min}}$. The evolution of the position of the interface along the major radius is shown in Fig. 2. Two different kinematic viscosities are considered for the droplet: $\upsilon_l = 0.05$ and $\upsilon_l = 0.1$. The kinematic viscosity of the vapor phase is set to be 0.3. As can be seen in Fig. 2, the droplet viscosity exerts an influence on the amplitude of the oscillation, but will not affect the oscillation period. The numerically predicted oscillation period is 2600, which agrees well with the analytical result $T_a \approx 2593.8$.

### B. Interface thickness and spurious velocity

It is well-known that the LB method is a diffuse interface method for modeling multiphase flows. In diffuse interface methods, the sharp fluid-fluid interface is replaced by a narrow layer in which the fluids mix [44]. In the literature, much research has shown that [17, 25], for LB simulations of dynamic multiphase flows, the width of the mixed layer (namely the interface thickness) should be around 4 ~ 5 lattices.

Through a simple algebraic procedure, the C-S equation of state can be non-dimensionalized as follows (see Appendix B for details):

$$p_{EOS} = 2.786 p_c \bar{\rho} \left[ \bar{T} \frac{1 + 0.13045\bar{\rho} + (0.13045\bar{\rho})^2 - (0.13045\bar{\rho})^3}{(1 - 0.13045\bar{\rho})^3} - 1.3829\bar{\rho} \right], \qquad (33)$$

where $\bar{\rho} = \rho/\rho_c$ and $\bar{T} = T/T_c$. According to Eqs. (22) and (33), the non-dimensional coexistence curve given by the Maxwell construction is dependent on $\bar{T}$. For the C-S equation of state, $\rho_c \approx 0.5218/b$. Hence the dimensional coexistence curve will be determined by both $b$ and $\bar{T}$. In addition, from Eq. (33) it can be seen that the magnitude of $p_{EOS}$ is related to $p_c$, which is given by $p_c = 0.070663 \, a/b^2$. Obviously, when $b$ and $\bar{T}$ are given, the parameter $a$ will determine the magnitude of $p_{EOS}$. In some of the existing multiphase LB models, the following equation of state is adopted [17]:

$$p = 4\beta'\rho(\rho-\rho_l)(\rho-\rho_g)\left[\rho-0.5(\rho_l+\rho_g)\right] - \beta'(\rho-\rho_l)^2(\rho-\rho_g)^2, \tag{34}$$

and it has been shown that the interface thickness in these models is proportional to $1/\sqrt{\beta'}$. Similarly, we believe that the interface thickness in the present pseudopotential model may be related to $1/\sqrt{a}$.

To numerically investigate the influence of the parameter $a$ on the interface thickness, three different values of $a$ are considered: $a = 1.0$, $0.5$, and $0.25$. The estimated interface thickness (in lattice units) is plotted in Fig. 3. From the figure we can see that, for a given $T/T_c$, the interface thickness obtained with $a = 0.25$ is larger than the interface thickness given by $a = 0.5$, which is in turn larger than that of $a = 1.0$. Moreover, it can be found that the results of the cases $a = 0.25$ and $0.5$ are about 1.9 and 1.4 times of that given by $a = 1.0$, respectively, which indicates that the interface thickness is approximately proportional to $1/\sqrt{a}$.

With the increase of the interface thickness, it is expected that the spurious currents will be reduced. To illustrate this point, the maximum magnitude of the spurious velocities of the three cases at $T/T_c = 0.5$ and $T/T_c = 0.55$ with $\tau_\upsilon = 0.8$ are listed in Table I, which shows that the spurious velocity can be reduced by a factor of $15 \sim 20$ from $a = 1.0$ to $0.25$. We therefore conclude that $a = 0.25$ is the best choice for the large-density-ratio regime ($T/T_c \leq 0.55$) since it gives an interface width of $4 \sim 5$ lattices in this regime and can greatly reduce the spurious currents as compared with $a = 1.0$.

Although the parameter $a$ has no effect on the solution of the Maxwell construction, it will affect the mechanical stability solution of the pseudopotential LB model because the potential $\psi$ varies with $a$, which can be seen from Eq. (24). To achieve thermodynamic consistency, the parameter $\sigma$ in the improved forcing scheme can be slightly changed with $a$. In the previous section, we have shown that $\sigma = 0.11$ ($\varepsilon = 1.76$) is used for $a = 0.5$. For $a = 0.25$, $\sigma$ can be chosen as $\sigma = 0.114$ ($\varepsilon = 1.824$). The solutions of these two cases at $\tau_\upsilon = 0.8$ are compared in Table II. It can be seen that both of them are in good agreement with the solution given by the Maxwell construction. Here it should also be pointed out that, for circular interfaces, the coexistence liquid and vapor densities will vary with the droplet size according to the Laplace's law [45]. To reduce the influence of droplet size, an alternative choice may be a piecewise equation of state, which can offer a separate control of $\partial p/\partial \rho$ in every single phase region and the mixed region [52].

### C. Droplet splashing on a thin liquid film

In this section, numerical simulations are carried out for the problem of a droplet with an initial velocity splashing on a thin liquid film. Actually, splashing can occur at widely different scales, from the astronomical scale when a comet impacts a planet to the microscopic scale in laboratory experiments [46-48]. The splashing of droplets on liquid/solid surfaces is a crucial event in a wide variety of phenomena in natural process and industrial applications, such as a raindrop splashing on the ground, the impact of a fuel droplet on the wall of a combustion chamber, and nano-printing using the laser induced forward transfer technique.

In our simulations, a two-dimensional planar droplet is considered and a grid size of $600 \times 250$ is adopted. The liquid film is placed at the bottom of the computational domain and its height is one-tenths of the entire domain height. The radius of the droplet is $R = 50$ and its impact velocity is $(v_x, v_y) = (0, -U)$, where $U = 0.125c$ ($c = \delta_t = 1$). The no-slip boundary condition is applied in the $y$-direction and the periodic condition is employed in the $x$-direction. The relaxation times $\tau_e$ and $\tau_\varsigma$ are chosen as $\tau_e^{-1} = \tau_\varsigma^{-1} = 0.8$. The reduced temperature is set to be $T/T_c = 0.5$ as it gives an equilibrium density ratio around 700, which is close to the water/air density ratio ($\approx 773$). The parameter $a$ is set to be 0.25 with $T_c \approx 0.0235$.

In almost all the existing studies of the pseudopotential LB model, the dynamic viscosity ratio $\mu_l/\mu_g = (\rho_l/\rho_g)/(\upsilon_g/\upsilon_l)$ is equal to the density ratio $\rho_l/\rho_g$ as in these studies the same relaxation time was used in the whole computational domain, which leads to $\upsilon_g/\upsilon_l = 1$. Here $\upsilon_g/\upsilon_l$ is the kinematic viscosity ratio between the vapor and liquid phases. Under such a condition, when the density ratio is around 1000, a very large dynamic viscosity ratio as well as a sharp change of the viscous stress tensor will be encountered in the interface, which will affect the numerical stability of the pseudopotential LB model and then make the liquid viscosity limited in a narrow range.

According to the molecular theory [49], the viscosity ratio is a function of the density ratio. When $\rho_l$ and $\rho_g$ are given, the viscosity ratio $\upsilon_g/\upsilon_l$ will be determined. Following the molecular theory, in the LB community Suryanarayanan *et al*. [50] have adopted a variable viscosity ratio for simulating dense gases. In the present work, in order to investigate the effect of the viscosity ratio on the pseudopotential LB model, we employ various values of $\upsilon_g/\upsilon_l$ for given $\rho_l$ and $\rho_g$, and it is found that a lower liquid kinematic viscosity can be gained with the increase of $\upsilon_g/\upsilon_l$. The lowest achievable liquid kinematic viscosity at $T/T_c = 0.5$ is plotted in Fig. 4 as a function of the ratio $\upsilon_g/\upsilon_l$. From the figure we can see that, for $\upsilon_g/\upsilon_l = 1$ the achievable lowest $\upsilon_l$ is about 0.075, while at $\upsilon_g/\upsilon_l = 20$ the liquid kinematic viscosity can be lowered to 0.009, and further to about 0.0039 when $\upsilon_g/\upsilon_l = 50$. For simplicity, the viscosity in simulations can be taken as $\upsilon(\rho) = \upsilon_l$ for $\rho \geq \rho_c$ and $\upsilon(\rho) = \upsilon_g$ for $\rho < \rho_c$.

With the above strategy, three different cases are considered for the present test ($T/T_c = 0.5$): $Re = 40$, 100, and 1000. The Reynolds number is defined as $Re = UD/\upsilon_l$, in which $D$ is the diameter of the impact droplet. The case $Re = 1000$ is realized by setting $\upsilon_g/\upsilon_l = 15$, which is the kinematic viscosity ratio of air to water at room temperature and normal atmospheric pressure. The Weber number $We = \rho_l U^2 D/\vartheta \approx 103$ (the surface tension $\vartheta$ is evaluated via the Laplace's law). The snapshots of the impingement process at $Re = 40$, 100, and 1000 are shown in Figs. 5, 6, and 7, respectively. The non-dimensional time is defined as $t^* = Ut/D$. From the figures we can see that, at $Re = 40$, the impact of the droplet will not result in splashing but an outward moving surface wave. With the increase of the Reynolds number, as can be seen in Fig. 6, a thin liquid sheet will be emitted after the impact of the droplet, which will grow into a crown propagating radially away from the droplet.

For larger Reynolds numbers ($Re = 1000$), a thinner liquid sheet will be formed in a small region located at the intersection between the droplet and the liquid layer. Then the sheet tilts upward and evolves into an almost vertical lamella whose end-rim is unstable and will eventually break-up into secondary droplets, which is an important phenomenon of droplet splashing and can be seen clearly in Fig. 7 at $t^* = 1.9$. All of these observations are in excellent agreement with the solution in Ref. [51], which was obtained using the volume-of-fluid method. For comparison, the density contours obtained with $a = 1.0$ at $Re = 1000$ and $t^* = 1.9$ are shown in Fig. 8 ($\upsilon_g/\upsilon_l$ is set to be 35 as the case is unstable at $\upsilon_g/\upsilon_l = 15$), from which some unphysical behavior can be clearly observed in comparison with Fig. 7(d).

In addition, previous research [46-48] has shown that the spread radius $r$ generally obeys the power law $r/D \approx C\sqrt{Ut/D}$ at short times after the impact. The sketch of the definition of the spread radius can be found in Ref. [46]. The coefficient $C$ is related to the setup of the problem. For three-dimensional modeling of droplet splashing and axisymmetric modeling of droplet splashing, Josseranda and Zaleskib found that $C \approx 1.1$ [46]. For two-dimensional modeling of droplet splashing, as can be seen in previous studies [17, 27, 48], the coefficient $C$ will be larger than $1.1$ owing to the fact that a two-dimensional planar droplet is a liquid cylinder rather than a spherical droplet in three-dimensional space. In Fig. 9, the predicted spread factors $r/D$ at $\text{Re} = 100$ and $1000$ are plotted as a function of the non-dimensional time $Ut/D$. As can be observed, there is no obvious dependence of the spread radius on the Reynolds number and the present numerical results are in overall accord with the prediction of the power law $r/D = 1.3\sqrt{Ut/D}$.

## V. CONCLUSIONS

In this paper, we have presented an improved MRT pseudopotential LB model via proposing an improved forcing scheme. Through numerical simulations of stationary droplet and droplet oscillation, the improved forcing scheme has been demonstrated to be capable of achieving both thermodynamic consistency and large density ratio in the MRT pseudopotential LB model. Subsequently, the influences of the parameter $a$ in the C-S equation of state on the interface thickness and the spurious velocity have been investigated. We found that the interface thickness is approximately proportional to $1/\sqrt{a}$. Meanwhile, $a = 0.25$ is found to be suitable for the large-density-ratio regime ($T/T_c \leq 0.55$) since it gives an interface thickness of $4 \sim 5$ lattices in this regime and can significantly reduce the spurious currents as compared with $a = 1.0$, which is widely used in previous studies.

Furthermore, the effect of the kinematic viscosity ratio $\upsilon_g/\upsilon_l$ has also been investigated. It is found that a lower liquid viscosity can be obtained in the pseudopotential LB model with the increase of $\upsilon_g/\upsilon_l$. With the above strategies, numerical simulations of two-dimensional droplet splashing on a thin liquid film have been successfully conducted at a density ratio larger than 500 with the Reynolds number from 40 to 1000. In our simulations, the dynamics of droplet splashing with increasing Reynolds number is correctly reproduced. The crown-like sheet and the formation of secondary droplets, which is an important phenomenon of droplet splashing, are well captured and the spread radius is found to obey the power law $r/D \approx 1.3\sqrt{Ut/D}$.

In summary, in the present study we have made an attempt to extend the pseudopotential LB model to the simulations of multiphase flows at large density ratio and relatively high Reynolds number. The related treatments can be summarized as follows. First, an improved forcing scheme is devised for the MRT pseudopotential LB model in order to achieve thermodynamic consistency and large density ratio. Second, a suitable value is chosen for the parameter $a$ in the C-S equation of state so as to obtain an interface thickness around 4 ~ 5 lattices in the large-density-ratio regime. Last but not least, an appropriate kinematic viscosity ratio is applied, which can be used to lower the liquid viscosity. These strategies (as a whole) may be useful for prompting the application of the pseudopotential LB model in multiphase flows at large density ratio and relatively high Reynolds number.

## ACKNOWLEDGMENTS

Support by the Engineering and Physical Sciences Research Council of the United Kingdom under Grant No. EP/I012605/1 is gratefully acknowledged.

## APPENDIX A: DERIVATION OF THE MECHANICAL STABILITY CONDITION

In this Appendix, the derivation of Eq. (21) is given. To start with, we can rewrite Eq. (20) as

$$P_n = \rho c_s^2 + \frac{Gc^2}{2}\psi^2 + \frac{Gc^4}{12}\left[\alpha\left(\frac{d\psi}{dn}\right)^2 + \beta\frac{\psi}{2}\frac{d}{d\psi}\left(\frac{d\psi}{dn}\right)^2\right], \qquad (A1)$$

using

$$\frac{1}{2}\frac{\mathrm{d}}{\mathrm{d}\psi}\left(\frac{\mathrm{d}\psi}{\mathrm{d}n}\right)^2 = \frac{\mathrm{d}^2\psi}{\mathrm{d}n^2}.\tag{A2}$$

By representing $(\mathrm{d}\psi/\mathrm{d}n)^2$ with $\varphi$, we can obtain

$$\alpha\varphi + \beta\frac{\psi}{2}\frac{\mathrm{d}\varphi}{\mathrm{d}\psi} = \frac{\beta}{2}\psi^{1+\varepsilon}\frac{\mathrm{d}}{\mathrm{d}\psi}\left(\psi^{-\varepsilon}\varphi\right) = \frac{\beta}{2}\frac{\psi^{1+\varepsilon}}{\psi'}\frac{\mathrm{d}}{\mathrm{d}\rho}\left(\psi^{-\varepsilon}\varphi\right),\tag{A3}$$

where $\varepsilon = -2\alpha/\beta$ and $\psi' = \mathrm{d}\psi/\mathrm{d}\rho$. Note that $\varphi = (\mathrm{d}\psi/\mathrm{d}n)^2 = \psi'^2(\mathrm{d}\rho/\mathrm{d}n)^2$. Then Eq. (A1) can be rewritten as

$$P_n = \rho c_s^2 + \frac{Gc^2}{2}\psi^2 + \frac{Gc^4\beta}{24}\frac{\psi^{1+\varepsilon}}{\psi'}\frac{\mathrm{d}}{\mathrm{d}\rho}\left[\frac{\psi'^2}{\psi^\varepsilon}\left(\frac{\mathrm{d}\rho}{\mathrm{d}n}\right)^2\right].\tag{A4}$$

According to Eq. (A4), we can obtain

$$\left(P_n - \rho c_s^2 - \frac{Gc^2}{2}\psi^2\right)\frac{24}{Gc^4\beta}\frac{\psi'}{\psi^{1+\varepsilon}} = \frac{\mathrm{d}}{\mathrm{d}\rho}\left[\frac{\psi'^2}{\psi^\varepsilon}\left(\frac{\mathrm{d}\rho}{\mathrm{d}n}\right)^2\right].\tag{A5}$$

Integrating Eq. (A5) leads to ($G$, $c$, and $\beta$ are constants)

$$\frac{24}{Gc^4\beta}\int_{\rho_g}^{\rho_l}\left(P_n - \rho c_s^2 - \frac{Gc^2}{2}\psi^2\right)\frac{\psi'}{\psi^{1+\varepsilon}}\mathrm{d}\rho = \int_{\rho_g}^{\rho_l}\mathrm{d}\left[\frac{\psi'^2}{\psi^\varepsilon}\left(\frac{\mathrm{d}\rho}{\mathrm{d}n}\right)^2\right],\tag{A6}$$

which gives

$$\frac{24}{Gc^4\beta}\int_{\rho_g}^{\rho_l}\left(P_n - \rho c_s^2 - \frac{Gc^2}{2}\psi^2\right)\frac{\psi'}{\psi^{1+\varepsilon}}\mathrm{d}\rho = \left.\frac{\psi'^2}{\psi^\varepsilon}\left(\frac{\mathrm{d}\rho}{\mathrm{d}n}\right)^2\right|_{\rho_g}^{\rho_l}.\tag{A7}$$

In every single phase far from the interface the pressure $P_n$ at equilibrium should satisfy

$$P_n(\rho_l) = \rho_l c_s^2 + \frac{Gc^2}{2}\psi(\rho_l)^2,\quad P_n(\rho_g) = \rho_g c_s^2 + \frac{Gc^2}{2}\psi(\rho_g)^2.\tag{A8}$$

Namely $(\mathrm{d}\rho/\mathrm{d}n)$ is zero in every single phase region, hence we have

$$\left.\frac{\psi'^2}{\psi^\varepsilon}\left(\frac{\mathrm{d}\rho}{\mathrm{d}n}\right)^2\right|_{\rho_g}^{\rho_l} = 0.\tag{A9}$$

From Eqs. (A7) and (A9), we can obtain

$$\int_{\rho_g}^{\rho_l}\left(P_n - \rho c_s^2 - \frac{Gc^2}{2}\psi^2\right)\frac{\psi'}{\psi^{1+\varepsilon}}\mathrm{d}\rho = 0.\tag{A10}$$

**APPENDIX B: NON- DIMENSIONALIZATION OF THE C-S EQUATION OF STATE**

The C-S equation of state can be non-dimensionalized via $\bar{\rho} = \rho/\rho_c$ and $\bar{T} = T/T_c$, in which $\rho_c$ and $T_c$ are the critical density and temperature, respectively. For the C-S equation of state, the critical density is given by $\rho_c \approx 0.5218/b$. Then $b\rho/4 = b\bar{\rho}\rho_c/4 \approx 0.13045\bar{\rho}$. Consequently, the C-S equation of state can be rewritten as

$$p_{\text{EOS}} = \rho \left[ RT \frac{1 + 0.13045\bar{\rho} + (0.13045\bar{\rho})^2 - (0.13045\bar{\rho})^3}{(1 - 0.13045\bar{\rho})^3} - a\rho \right]. \tag{B1}$$

Since $a = 0.4963 R^2 T_c^2/p_c$ and $b = 0.18727 RT_c/p_c$, $a\rho \approx 0.5218\bar{\rho} a/b = 1.3829\bar{\rho} RT_c$. With this result, Eq. (B1) becomes

$$p_{\text{EOS}} = \rho RT_c \left[ \bar{T} \frac{1 + 0.13045\bar{\rho} + (0.13045\bar{\rho})^2 - (0.13045\bar{\rho})^3}{(1 - 0.13045\bar{\rho})^3} - 1.3829\bar{\rho} \right]. \tag{B2}$$

Meanwhile, $\rho RT_c \approx 0.5218 \bar{\rho} RT_c/b \approx 2.786 \bar{\rho} p_c$. Hence the non-dimensionalized C-S equation of state is given by

$$p_{\text{EOS}} = 2.786 p_c \bar{\rho} \left[ \bar{T} \frac{1 + 0.13045\bar{\rho} + (0.13045\bar{\rho})^2 - (0.13045\bar{\rho})^3}{(1 - 0.13045\bar{\rho})^3} - 1.3829\bar{\rho} \right]. \tag{B3}$$

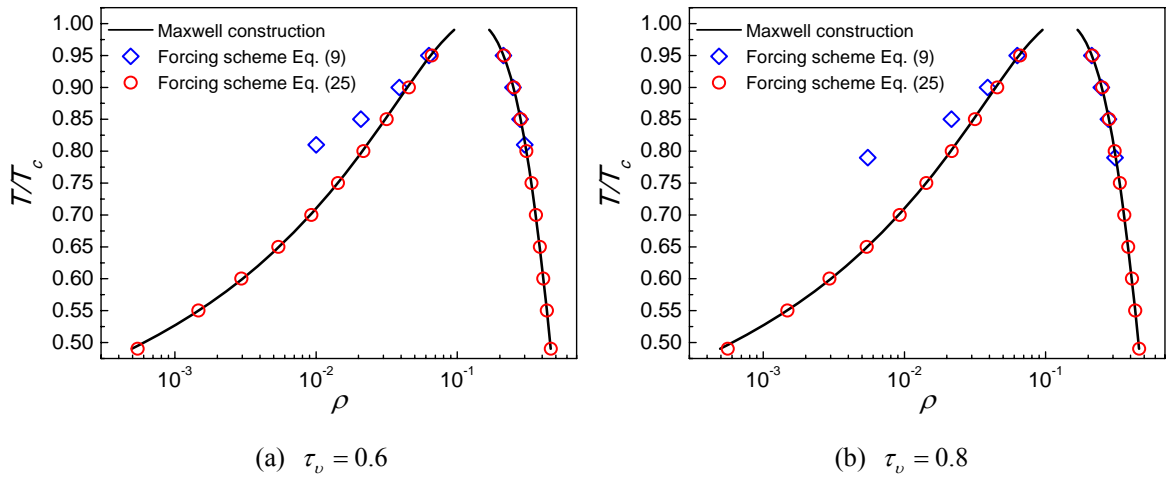

FIG. 1. (Color online) Comparison of the numerical coexistence curves predicted by the original and improved forcing schemes with the coexistence curves given by the Maxwell construction.

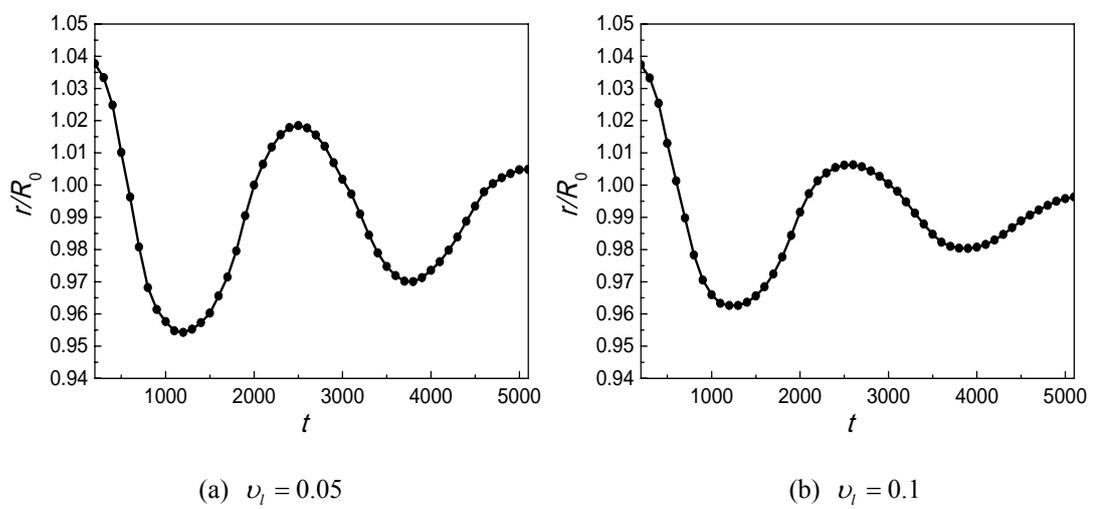

(a) $\upsilon_l = 0.05$        (b) $\upsilon_l = 0.1$

FIG. 2. Oscillation of an elliptic droplet at $T/T_c = 0.5$ ($\rho_l/\rho_g \sim 700$) with different liquid viscosities.

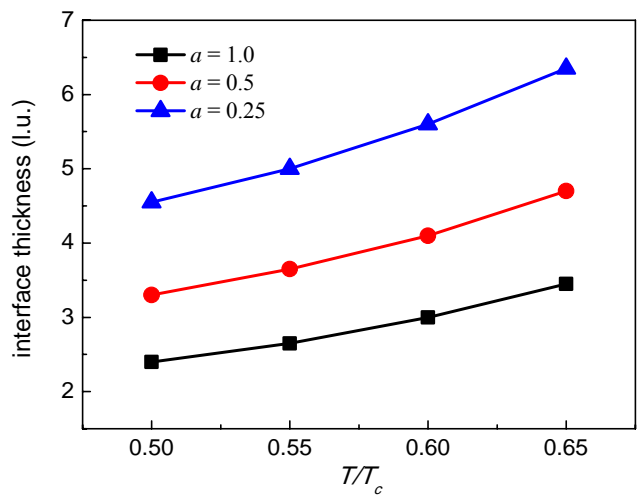

FIG. 3. (Color online) Comparison of the interface thickness obtained via different $a$ in the C-S equation of state. The l.u. represents lattice unit.

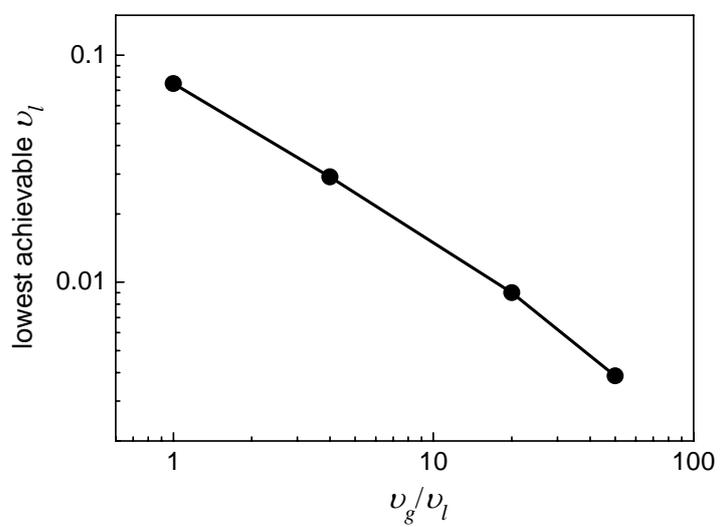

FIG. 4. Simulation of droplet splashing on a thin liquid film at $T/T_c = 0.5$: the lowest achievable liquid kinematic viscosity $\upsilon_l$ as a function of the kinematic viscosity ratio $\upsilon_g/\upsilon_l$.

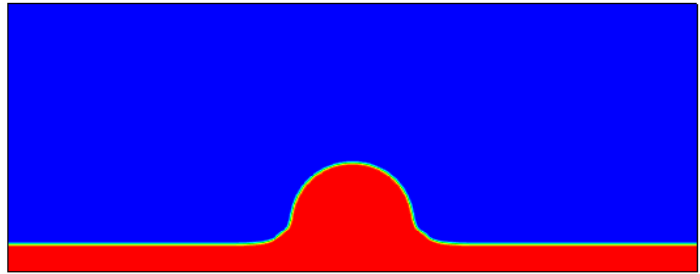

(a) $t^* = 0.25$

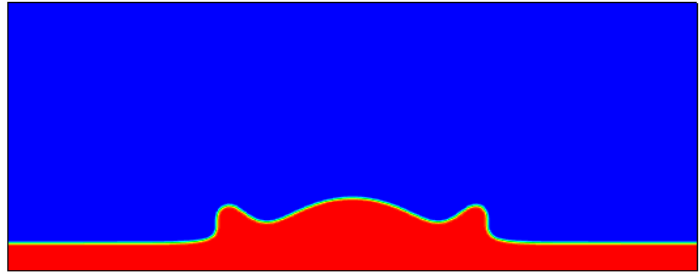

(b) $t^* = 0.75$

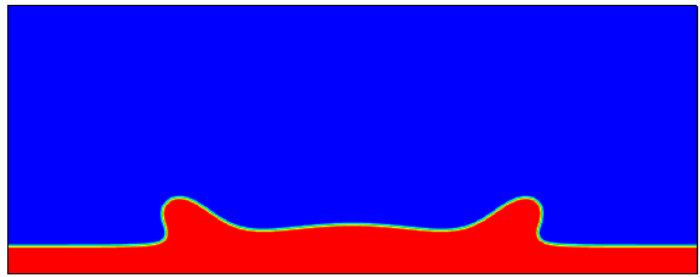

(c) $t^* = 1.5$

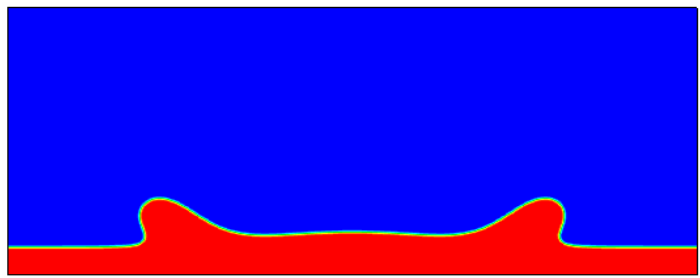

(d) $t^* = 1.9$

FIG. 5. (Color online) Snapshots of the impingement process at $T/T_c = 0.5$ and $\mathrm{Re} = 40$.

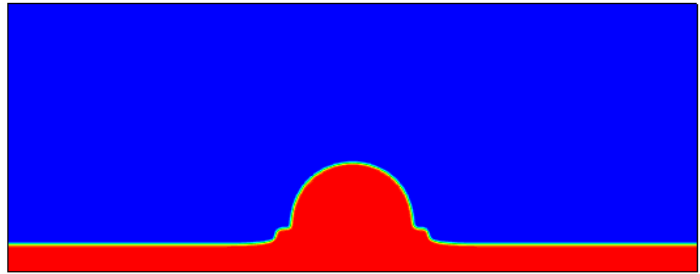
(a) $t^* = 0.25$

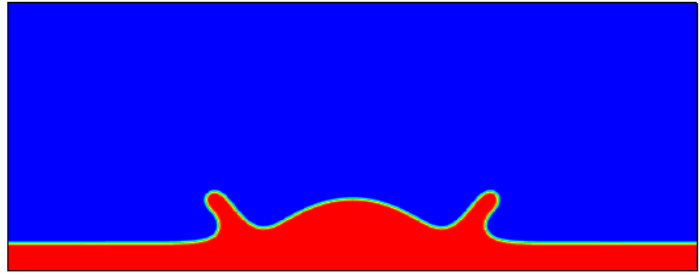
(b) $t^* = 0.75$

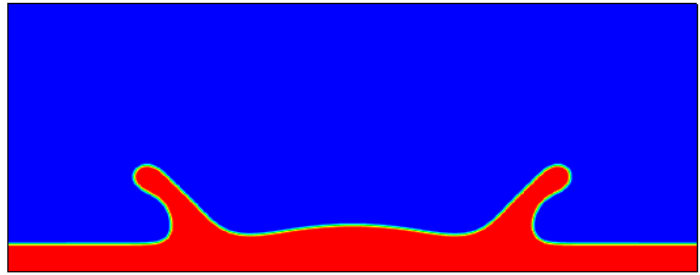
(c) $t^* = 1.5$

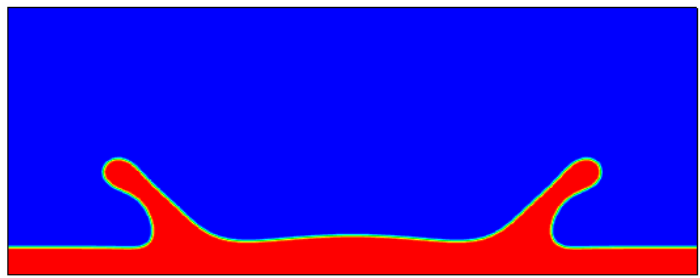
(d) $t^* = 1.9$

FIG. 6. (Color online) Snapshots of the impingement process at $T/T_c = 0.5$ and $Re = 100$.

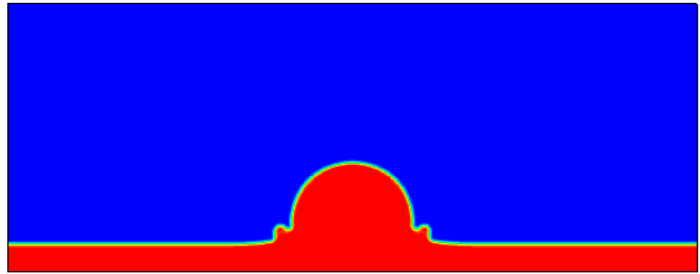

(a) $t^* = 0.25$

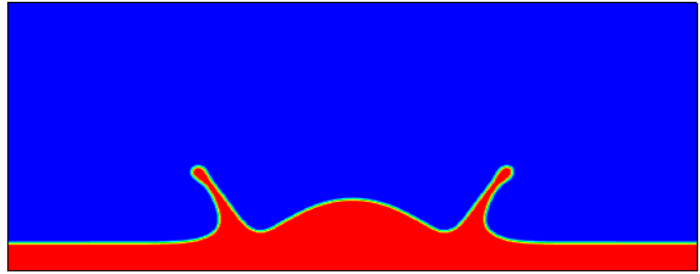

(b) $t^* = 0.75$

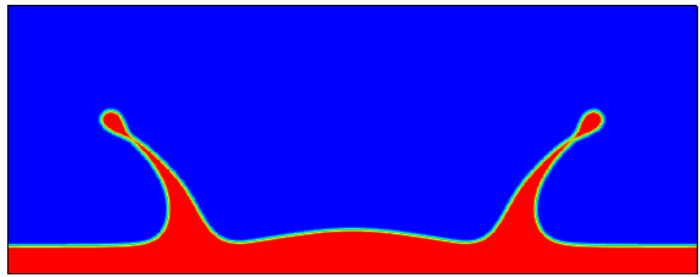

(c) $t^* = 1.5$

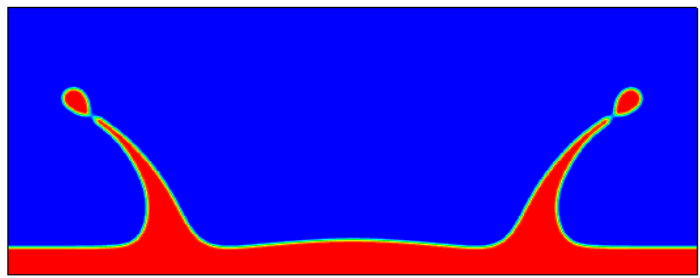

(d) $t^* = 1.9$

FIG. 7. (Color online) Snapshots of the impingement process at $T/T_c = 0.5$ and $\mathrm{Re} = 1000$.

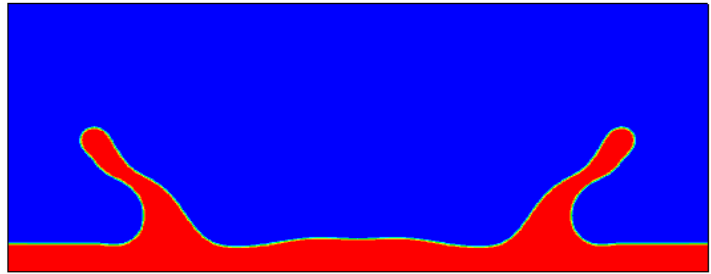

FIG. 8. (Color online) Density contours obtained with $a = 1.0$ at $\text{Re} = 1000$ and $t^* = 1.9$ ($T/T_c = 0.5$).

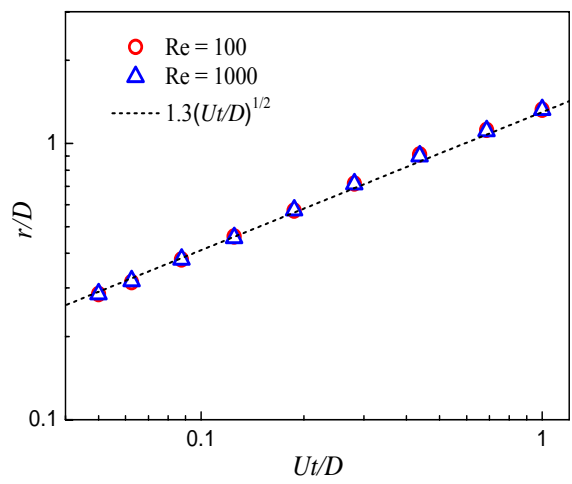

FIG. 9. (Color online) The predicted spread radius at Re = 100 and 1000 as a function of the non-dimensional time.

Table I. Comparison of the maximum magnitude of the spurious velocities obtained by different $a$.

| $T/T_c$ | density ratio | $a = 1.0$ | $a = 0.5$ | $a = 0.25$ |
|---|---|---|---|---|
| 0.55 | 293.2 | 0.0399 | 0.00786 | 0.00256 |
| 0.50 | 750.8 | 0.0733 | 0.0136 | 0.00390 |

Table II. Comparison of the densities $\rho_l$ and $\rho_g$ obtained by $a = 0.5$ and $0.25$.

| $T/T_c$ | $\rho_l/\rho_g$ | | |
|---|---|---|---|
| | numerical ($a = 0.5$) | numerical ($a = 0.25$) | Maxwell construction |
| 0.60 | 0.4077/0.00298 | 0.4079/0.00306 | 0.407/0.00300 |
| 0.55 | 0.4317/0.001484 | 0.4318/0.001484 | 0.431/0.00147 |
| 0.50 | 0.4559/0.000667 | 0.4547/0.000639 | 0.455/0.000606 |